\documentclass[prc,aps,amssymb,twocolumn,showpacs]{revtex4-1} 
\usepackage{graphicx}
\usepackage{hyperref}
\usepackage{natbib}

\begin{document}
\pagebreak
\title{Relation Between Wigner Energy and Proton-Neutron  Pairing}
\author{I. Bentley$^{1,2}$}
\author{S. Frauendorf$^{1}$}
\affiliation{$^1$Dept. of Physics, University of Notre Dame, Notre Dame, IN
46556}
\affiliation{$^2$Dept. of Chemistry and Physics, Saint Mary's College, Notre
Dame, IN
46556}

\date{\today}

\begin{abstract} 
The linear term proportional to $|N-Z|$ in the nuclear symmetry energy (Wigner
energy)
is obtained in a model that uses isovector pairing on single particle levels
from a deformed potential combined with a $\vec T^2$ interaction. 
The pairing correlations are calculated by numerical diagonalization of the
pairing Hamiltonian acting on the six or seven levels nearest the $N=Z$ Fermi surface.
The experimental binding energies of nuclei with $N\approx Z$
are well reproduced. The Wigner energy emerges as a consequence of restoring
isospin symmetry. We have found the Wigner energy to be insensitive to the presence of moderate isoscalar pair correlations.
\end{abstract}


\maketitle

\section{Introduction}
The nuclear ground state energy, $E(N,Z)$, as a function of the proton number
($Z$)
and neutron number ($N$) or atomic mass number ($A=N+Z$) is very well described 
by the celebrated empirical mass
formula (see e.g. \citep{MS82}):
\begin{equation}
E(N,Z)=E_V+E_S+E_C+E_A+E_W+E_P+E_{SHELL}.
\end{equation}
The various terms have a clear physical meaning. The volume term, $E_V=-a_V A$,
describes the constant binding energy of
a nucleon in  saturated nuclear matter. The surface energy, $E_S=a_SA^{2/3}$,
accounts for the lack of neighbors in the surface.  The term,
$E_C=a_CZ^2/A^{1/3}$, is the electrostatic Coulomb energy. The (a-)symmetry 
energy, $E_A=a_A(N-Z)^2/A$, consists of two approximately equal contributions. 
The \textquotedblleft kinetic\textquotedblright part accounts for the Pauli principle, which requires the
nucleons
to occupy higher single particle
levels with increasing asymmetry $|N-Z|$.
The \textquotedblleft interaction\textquotedblright part originates from the difference between the
proton-proton, neutron-neutron and proton-neutron interactions. The pairing energy ($E_P$)
describes the energy gain by forming
pairs of protons or neutrons.
The shell energy ($E_{SHELL}$) is a manifestation of  the level bunching around 
the Fermi level. 
The term, $E_W=a_W|N-Z|/A$, is called the Wigner energy, because Wigner
\citep{Wi37}
gave a first interpretation in terms of his super multiplet theory. However, its
physical origin has been the subject of a long debate, which has been recently
reviewed 
by \citep{Ne09}. Modern mean field approaches reproduce the  ground
state energies very well, except the Wigner energy, which has to be added as an
ad-hoc phenomenological term (see e.g. \citep{HFB21}). This means that the
physics behind the Wigner energy is not taken into account by present mean field
theories. 

In this letter we demonstrate that the Wigner energy is obtained,
without introducing any new parameters, by including the isovector
proton-neutron pair
correlations determined by numerical diagonalization of
an isorotational invariant  pairing Hamiltonian.  

Experimentally, the coefficients $a_A$ and $a_W$ are not very different. 
 As the ground state isospin ($T$) of most 
nuclei is equal to their isospin projection $(T_z=\frac{N-Z}{2})$. The sum of
the symmetry and Wigner energies
is approximately proportional to $T(T+1)$. The $T$ - dependence is suggestive,
because the isospin operators obey
 the same SU$_2$ algebra as the angular momentum operators. Spontaneous breaking
of the rotational symmetry
 by the deformed mean field leads to the appearance of rotational bands. The
energies of the rotational levels are 
 proportional to $I(I+1)$, with $I$ being the angular momentum. The analogy
between nuclear spin and isospin led Frauendorf and Sheikh \citep{FS99,FS00} to
suggest that the $T(T+1)$ dependence of the ground state energy is a
manifestation of an isorotational band.

The band appears because the isovector pair field, which is
a vector, spontaneously breaks rotational
 symmetry in isospace. Glowacz, Satula and Wyss  discussed the analogy of the cranking
model in isospace \citep{SW02,Gl04}.
  In the limit of strong symmetry breaking,  simply the isorotational energy
$T(T+1)/2\Theta$
 is added to the intrinsic energy of the symmetry breaking mean field, the
orientation of
which can be taken such that the proton-neutron
 pair field is zero \citep{FS99,FS00}. Afanasjev {\em et al.} \citep{Le03,AF05, ASA07} successfully used 
 this simple limit to interpret the excitation 
 spectra of nuclei with $N\approx Z$. 

 In a series of papers, J\"anecke and coworkers \citep{Ja05} and earlier work
cited therein, \citep{JO02} and \citep{JO05},  demonstrated that the
 global $N-Z$ dependence of the binding energies, including the Wigner term and
the 
 inversion of the $T=0$ and $T=1$ states in odd-odd $N=Z$ nuclei with $A>40$,  can be well
understood in terms
 of the competition between the familiar pair gap $\Delta$ and a symmetry energy
term
 of the form $T(T+1)$.  

Applying the Mean Field and Random
Phase Approximation to an isorotational invariant isovector pairing interaction,
 Neerg\aa rd has reproduced the experimental observation $a_A \approx a_W$ 
\citep{Ne02,Ne03,Ne09}. 
 The virtue of such an approach is
that   the Wigner energy appears without introducing any new parameter, because 
the strength of the  proton-neutron pair correlation is fixed by the
isorotational invariance of the isovector pairing Hamiltonian.  

However, this approach only works
well when sufficiently far from the critical coupling strength for the appearance of the isovector pair field.
This cannot be  expected to be always the case in the medium
 mass nuclides, where the Wigner energy plays an important role. In order to avoid these
 problems, in this paper we treat the pair correlation into account by 
 numerically diagonalizing the isovector pair Hamiltonian within a configuration
space spanned by seven single particle levels nearest the Fermi surface. We demonstrate that
the detailed values of the Wigner energy depend on the level spacing at the Fermi surface, and that its variations with    
particle number can be reproduced using single particle energies of the Nilsson potential.  In addition, we
test  the robustness of the results with respect to presence of isoscalar pair correlations.

Section \ref{sec:exp} presents the separation of the Coulomb energy from experimental total binding energies
and describes how the experimental values of the Wigner energy,    
symmetry energy, and even-even odd-odd pair gaps are derived. The model is presented in 
section \ref{sec:MO} and its parameters are fixed in section \ref{sec:PA}. Section \ref{sec:res}  contains
the results for a pure isovector pair interaction. The consequences of an additional isoscalar pair interaction are
discussed in section \ref{sec:IVIS}. The consequences of using a small number of single particle states
when calculating pair correlations are discussed in section \ref{sec:FR}. 

\section{Extraction of the relevant experimental data}\label{sec:exp}
\subsection{Coulomb energy} \label{sec:CF}
As a starting point we assume that the isospin mixing caused by the Coulomb
interaction can be neglected. Ref. \citep{PhysRevLett.103.012502} 
estimated the admixture of
components with $T>T_z$ to the ground state to be of the order of \mbox{0.9 \%}
for
$A\sim70$.  With this assumption, the Coulomb energy can be separated from the
energy caused by the strong interaction. 
Following  \citep{Ja05} we subtract the Coulomb energy from the experimental
energies and compare the resulting energies with our model.
Because the mass tables have been revised meanwhile, a we repeated the extraction
of the strong interaction part of the binding energies.

The expression for the  Coulomb energy given in Ref. \citep{NR95} is adopted:
\footnotesize
\begin{equation}\label{eq:EC}
 E_C=\frac{3}{5}\frac{q^2}{4 \pi\epsilon_0 r_0} \frac{Z^2}{A^{1/3}}
\Bigg(1-\frac{5}{6} \bigg(\frac{d\pi}{r_0 A^{1/3}}\bigg)^2-5\bigg(\frac{3}{16\pi
Z}\bigg)^{2/3}\Bigg),
\end{equation}
\normalsize
where the two unknowns are the equivalent radius ($r_0$) and diffuseness ($d$).

This expression for Coulomb energy begins with the approximation that the
nucleus is a homogeneously charged sphere. The first correction 
takes into account diffuseness of the nuclear surface. The second is the
exchange correction which is necessary because protons obey the Pauli principle
and wave-functions cannot completely overlap (e.g. \citep{BM99}).

Using the finite difference approximation evaluated at an average value of
proton number ($\bar{Z}$), a more useful expression involving energy differences
results:
\footnotesize
\begin{equation}
\label{Cfitused}
 \frac{\partial E_C}{\partial Z} \bigg\vert _{Z=\bar{Z}} \approxeq \frac{\Delta
E_C}{\Delta Z}  \bigg\vert _{Z=\bar{Z}}
\end{equation}
\begin{displaymath}
=\frac{6}{5}\frac{q^2}{4 \pi\epsilon_0 r_0}
\frac{\bar{Z}}{A^{1/3}}\Bigg(1-\frac{5}{6} \bigg(\frac{d\pi}{r_0
A^{1/3}}\bigg)^2-\frac{10}{3}\bigg(\frac{3}{16\pi \bar{Z}}\bigg)^{2/3} \Bigg).
\end{displaymath}

\normalsize

\begin{figure}
  \begin{center}
    \includegraphics[width=8.7cm]{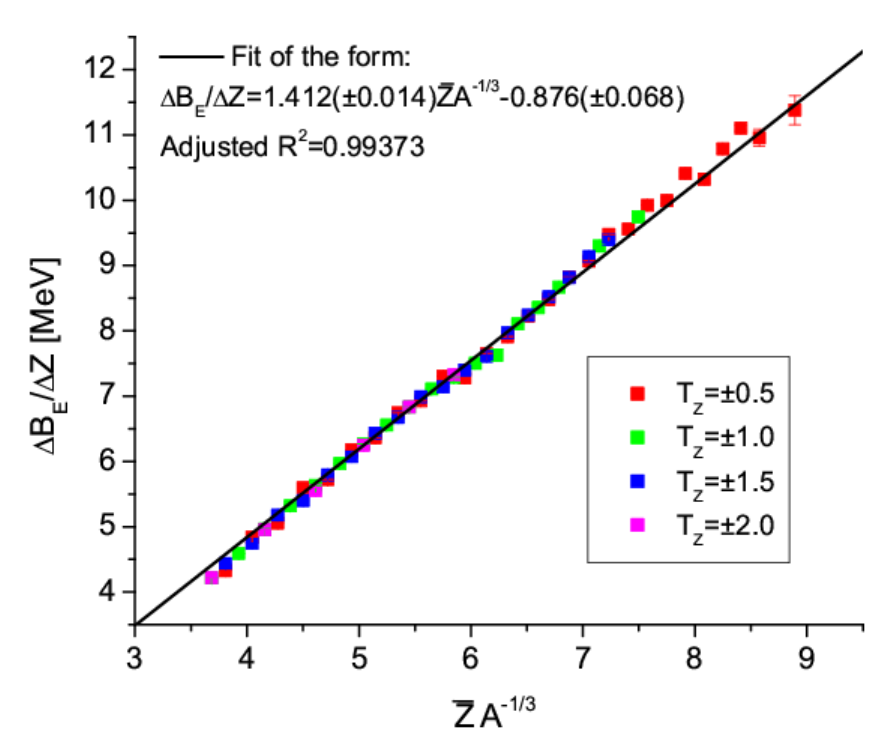}
    \caption{(Color on-line) Linear fit to experimental binding energy
differences of mirror nuclei plotted as a function of $\bar{Z} A^{-1/3}$. The
color indicates the isospin of the pair of mirror nuclei used.} 
    \label{fig:Coulomb}
  \end{center}
\end{figure}

For mirror nuclei, $\bar{Z}/A=1/2$ the two corrections, which depend on
powers of $\bar{Z}/A$ become constants. Experimental binding energies for
69 pairs of mirror nuclei  in the region $20\leq A \leq 100$ are found in the
2012 Atomic Mass
Evaluation (AME) \citep{Au12}. The fit shown in Figure \ref{fig:Coulomb} 
determines  the two unknowns  $r_0=1.224 fm$ and $d=0.281 fm$.
The diffuseness is not consistent with other estimates
because other higher order contributions to the Coulomb energy have been
combined \citep{NR95}.

The atomic binding energies used in these calculations also contain a small
contribution accounting for the binding of the electrons, which has been found to
be \citep{Lu03}:

\begin{equation}
 B_{el}(Z)=14.4381 Z^{2.39}+1.555 \times 10^{-6} Z^{5.35} eV.
\end{equation}
Each of the comparisons discussed here involve differences between neighboring
nuclei. For
these differences  term contributes at most about 15 keV. Nonetheless these
contributions are taken into account.

Applying the finite difference approximation to the  pairs of mirror nuclei
reduces Eqn. (\ref{eq:EC}) to two terms  because
$\bar{Z}=A/2$.  It can be fit with a root mean squared deviation of 104 keV as shown in Figure \ref{fig:Coulomb}, with
\begin{equation}
\label{eqn:DECMSE}
\frac{\Delta E_{C}}{\Delta Z} = 0.706 (\pm0.007) A^{2/3}-0.876 (\pm0.068) [MeV],
\end{equation}
which is comparable to previous fits (cf.  e.g.
\citep{Ja67}, \citep{Vo00}, \citep{DV07}). The $A$ -independent correction is determined by the radius, which was taken from fit of the slope. 
The remaining term depends on the diffuseness,
which was adjusted. The resulting expression for the differences of the Coulomb energy, in steps of $\Delta Z=2$, within 
an isobaric chain was used to calculate 
the differences between the strong interaction energies:
\small
\begin{displaymath}
E_S(Z+1,A)-E_S(Z-1,A)=E_{Exp}(Z+1,A)-E_{Exp}(Z-1,A)
\end{displaymath}
\begin{displaymath}
-2 \bigg[ 1.412 (\pm0.014) \frac{Z}{A^{1/3}}-0.610 (\pm0.048) \frac{Z}{A} 
\end{displaymath}
\begin{equation}
-0.719 (\pm0.007) \frac{Z^{1/3}}{A^{1/3}} \bigg] [MeV],
\end{equation}
\normalsize
which are needed in the expressions given in the next section.

The  uncertainties of our fit are given in parenthesis. They are propagated together with 
the quoted errors of the experimental binding energies to estimate the total error of the \textquotedblleft experimental\textquotedblright quantities shown in 
the following figures.  

\subsection{Experimental isorotational bands}\label{sec:IRB}

In accordance with the concept of isorotational bands, we write the energy of an
isobaric chain, with constant $A$, in the form:
\begin{equation}
E(N,Z)=E_{int}+\frac{T(T+X)}{2\Theta},~~T=\vert T_z\vert,
\end{equation}
where $E_{int}$ is the energy of the intrinsic ($N=Z$) configuration. As
discussed in \citep{FS99,FS00},
the  Bardeen-Cooper-Schrieffer (BCS) ground state without proton-neutron pairs is a legitimate
intrinsic state.
It is a mixture of only even  $N$ and $Z$, which implies that $T_z$ must be even
if $A/2$ is even or $T_z$ must be odd if $A/2$ is odd. Hence, the ground state
isorotational bands
of even-even nuclei are composed of even values of $T=T_z$ if $A/2$ is even and
odd values of $T=T_z$ if $A/2$ is odd. 

The term $1/2\Theta$ is a combination of  the
coefficient $a_S$ of the the symmetry energy and a contribution from the shell
energy $E_{SHELL}(Z,N)$,
which depends on $T_z$. 
Likewise, $X/\Theta$ is a combination related to  the coefficient $a_W$ of the
Wigner
energy, which also contains a contribution from $E_{SHELL}$. 
We introduce the experimental isorotational frequency
\begin{equation}\label{Eom}
\omega(T+1)=\frac{E(T+2)-E(T)}{2}=\frac{T+1+X}{\Theta}.
\end{equation}

The slope and intercept with the $\omega$-axis determine
$1/\Theta$ and $X$.
 We take the experimental ground state energies of the three nuclei with $T_z=$0, 2,
4 if $A/2$ is even
and with $T_z=$1, 3, 5  if $A/2$ is odd and calculate two points of $\omega(T)$ be
means of
Eqn. \ref{Eom}. 
 Note that this is just a recombination of the experimental ground state
energies, which 
 aims at exposing the  Wigner energy. 
 The explicit expressions for the experimental $X$ are
\footnotesize
\begin{equation}
\label{eqn:XE}
X_{E}(A)= \Bigg(\frac{6 E_S(T_z=0)- 8 E_S(T_z=2)+ 2 E_S(T_z=4)}{-E_S(T_z=0)+2
E_S(T_z=2)-E_S(T_z=4)} \Bigg),
\end{equation}
\normalsize
for even values of $T_z$  and
\footnotesize
\begin{equation}
\label{eqn:XO}
X_{O}(A)=  \Bigg(\frac{8 E_S(T_z=1)- 12 E_S(T_z=3)+ 4 E_S(T_z=5)}{-E_S(T_z=1)+2
E_S(T_z=3)-E_S(T_z=5)} \Bigg), 
\end{equation}
\normalsize
for odd $T_z$.

\begin{figure}
  \begin{center}
    \includegraphics[width=8.7cm]{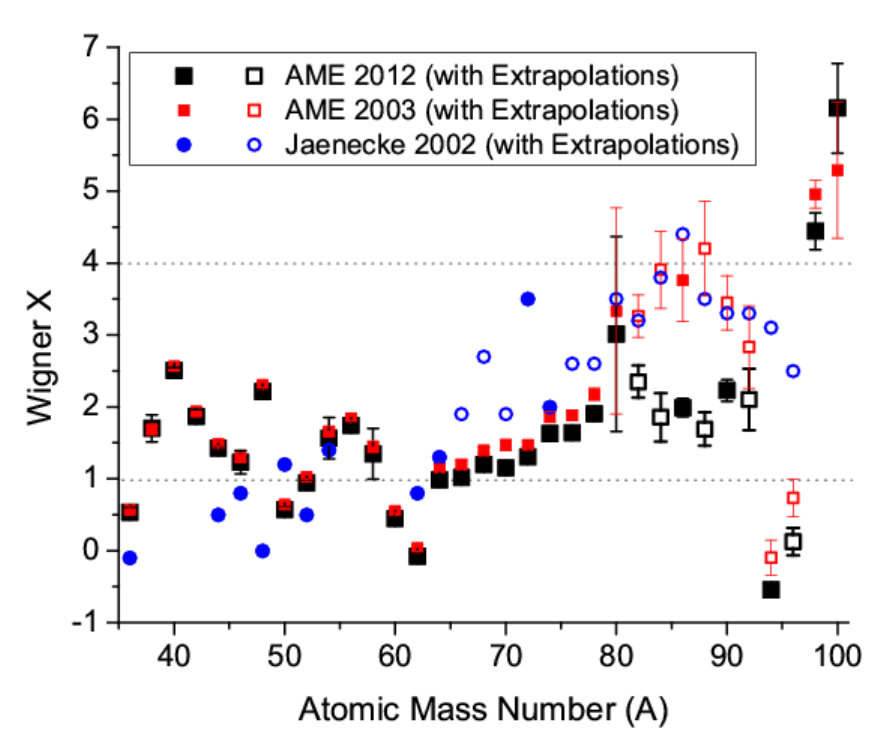}
    \caption{(Color on-line) The experimental Wigner X, derived from
(\ref{eqn:XE}) and (\ref{eqn:XO}). The isobaric chains were evaluated
using
experimental data from \citep{Au03} and \citep{Au12} with the Coulomb
contributions removed. The
$X$ values from
J\"{a}necke et al. have been included from a comparable Figure in \citep{JO02}.
Open symbols indicate values with at least one binding energy from an
extrapolation. If not visible, the error bars are smaller than the size of the symbols.} 
    \label{fig:Fig3.12}
  \end{center}
\end{figure}

Figure \ref{fig:Fig3.12}
 shows the
 experimental values of $X$, where the experimental binding energies 
 are taken from the most recent mass evaluation from \citep{Au12}.
The small error bars for the lower mass region are primarily caused by the
uncertainty in the spherical Coulomb energy fit. The large error bars in the
$A>80$ region are
mainly caused by the error in binding energy of the nucleus nearest to or at
$N=Z$.

A first determination of $X$ based on the 2003 AME \citep{Au03}
generally
reproduced the features described by J\"{a}necke et al. including an
apparent shift from $X\approx1$ to $X\approx4$ for $A>80$. However, one new
feature occurred near $A=92$ where $X\approx2$. A new observation
 is that as $A$ approaches a doubly magic nucleus, the value of $X$
appears to decrease then increase. 

J\"{a}necke suggested that the $A\approx80$ region, the large values of $X$ 
might be caused in part by the substantial uncertainties in the masses used \citep{JO02}. 
This speculation appeared to be in agreement with the reevaluation  of $X$ based
on the 2012 AME \citep{Au12}, upon which this paper is based.  There are several changes 
 in the $A=80-90$ region, which come from a combination of new
mass measurements, specifically for ($^{86}$Mo , and $^{90}$Ru) and
new extrapolations ($^{82}$Zr, $^{84}$Mo, $^{88}$Ru, and $^{92}$Pd) \citep{Au12}.
The systematic nature of the reduction of the value of $X$ in the $A=82-92$
chains is a result of the fact that the binding energy of the $T_Z=0,1$ nuclides  has
decreased by approximately 0.5 to 1 MeV \citep{Au12}. The lowering of these points
leads to smaller values of $X$.

Further changes  of the earlier evaluations result from new masses at $T_z=5$
($^{22}$C, $^{26}$O, $^{34}$Mg, and $^{38}$Si) and $N=Z$ ($^{100}$Sn) which have
changed by a few hundred keV or more. Elsewhere the difference in X between AME
2003 and 2012, results from the different Coulomb fits for the two data sets.
The results based on the 2012 masses have an average value of $X=1.64$ for
$24\leq A \leq100$.

\section{The Model}\label{sec:MO}

A monopole isovector pairing Hamiltonian is used to describe the
pair correlated ground state, 

\begin{eqnarray}
\label{eqn:IVPH}
 H_V=\sum_{k} \epsilon_{k} \hat{N}_k  - G_V
\sum_{kk', \tau} \hat{P}^+_{k,\tau}  \hat{P}_{k',\tau}+C\vec T\cdot\vec T,\\
\hat{N}_{k}=\hat{p}^+_{k}\hat{p}_{{k}}+\hat{p}^+_{\bar k}\hat{p}_{\bar{k}}+
\hat{n}^+_{k}\hat{n}_{{k}}+\hat{n}^+_{\bar k}\hat{n}_{\bar{k}},\\
 \hat{P}^+_{k,0}=\frac{1}{\sqrt{2}}\Big(\hat{n}^+_{k}\hat{p}^+_{\bar{k}}+\hat{p}
^+_{k}\hat{n}^+_{\bar{k}}\Big),\\
 \hat{P}^+_{k,-1}=\hat{p}^+_{k}\hat{p}^+_{\bar{k}},  \textrm{ and }
\hat{P}^+_{k,1}=\hat{n}^+_{k}\hat{n}^+_{\bar{k}},
\end{eqnarray}
where $\hat{p}^+_{k}$ and  $\hat{n}^+_{k}$ create a proton and a neutron,
respectively,
  on the level $k$, and $\bar k$ denotes the time reversed state of $k$.
  Identical single particle energies $\epsilon$ are used for protons and neutrons,
  which are derived from the Nilsson potential as described in section \ref{sec:PA}.
This Hamiltonian is invariant under rotations in isospace, i.e. it
conserves isospin. 

The many body  problem is solved  via matrix diagonalization. 
The space comprises the  single particle configurations
that are   generated from the lowest configuration with  $T=T_z$ by multiple application
of the interaction. The subsequent applications are carried out by a 
computer code, which stops when no new configurations are generated.
 As the dimension of the configuration space  grows quickly with
number of active single particle levels, it is assumed that the pair correlations 
are restricted  to the configurations within  the set of seven levels centered about  
Fermi level   $\epsilon_{k=N}$ of the $N=Z$ nuclide within the considered isobaric chain.   
All levels $\epsilon_{k<N-3}$ are assumed to be occupied and all levels  $\epsilon_{k>N+3}$
to be free.
Since the matrix is constructed by successive application of the isospin conserving 
pairing interaction onto the uncorrelated ground state, the configuration space contains only states with $T=T_z$. 
The isobaric chains  $T_z$=0, 2, 4 are studied if $A/2$ is even  or $T_z$=1, 3, 5 if  $A/2$ is odd.
The respective dimensions are 3647, 1890,  210 or  3647, 1001, 70. 
 In the case of odd-odd $N=Z$ nuclei, the lowest $T=1$ energy is equal to the energy of the $T=T_z=1$
isobar. As a test of the code, we also generated the configuration matrix for the odd-odd nucleus by
starting from the configuration with the odd proton-neutron pair on the Fermi level
in the $T=1$ state and diagonalized it. As it  has to be, the two energies agreed.    
The lowest $T=0$ state in odd-odd $N=Z$ nuclei was obtained by 
generating the configuration matrix 
starting from the configuration with the odd proton-neutron pair on the Fermi level
in the $T=0$ state and diagonalizing it. This is equivalent with blocking the Fermi level 
from the correlations because the isovector interaction cannot scatter the isoscalar pair onto other levels
nor scatter isovector pairs onto the Fermi level.

The diagonalization was carried out disregarding  the levels $\epsilon_{k<N-3}$ and $\epsilon_{k>N+3}$ .
To the energies resulting from the diagonalization  the sum of the
single particle energies for all occupied proton and neutron levels below the
seven-level window was added. As discussed in the next section, this ensures that
the shell correction to the binding energies is properly taken into account.

As suggested by Neerg\aa rd \citep{Ne09}, the term $C\vec T\cdot\vec T$ is a simple way to take into account the isospin
dependence 
of the single particle levels. The relation between the isospin dependence of
the 
nuclear potential and the  \textquotedblleft interaction\textquotedblright part of the symmetry energy has been
discussed 
by Bohr and Mottelson \cite{BM99}. It needs to be added, because we carry out
the 
diagonalization of the pairing Hamiltonian for a fixed set of single particle
levels along an 
isobaric chain. This means that only the \textquotedblleft kinetic\textquotedblright part of the symmetry energy
is taken into account.
The difference between the proton and neutron nuclear potentials generates an
orientation in isospace. Hence it must be included in the isorotational
energy. 
It appears in a natural way if one carries out \textquotedblleft isocranking\textquotedblright about the z-axis,
which is
just the standard procedure of fixing $\langle N \rangle$ and $\langle Z
\rangle$ in 
self-consistent Hartree-Fock-Bogoliubov calculations.

\section{Determination of the model parameters}\label{sec:PA}

The  single particle energies
are calculated by means of the Micro-Macro method using a Nilsson Hamiltonian as
described in Ref. \citep{Fr00}.
For each nucleus the equilibrium deformation has been calculated. In these
calculations BCS pairing was used
with $\Delta_Z=13.4 [MeV]/A^{1/2}$ and $\Delta_N=12.8 [MeV]/A^{1/2}$ as suggested in
\citep{MN92}.  Ref. \citep{BF11} discusses this procedure of determining the equilibrium deformation called 
AutoTAC in more detail.
The resulting deformations  are comparable with those from Ref. \citep{MN95}.
 The single particle  energies 
used in the diagonalization of the pairing Hamiltonian are  
taken as the average of the proton and neutron energies calculated 
by the Nilsson model at equilibrium deformation.

The use of the averaged energies is justified as follows. The premise of our model is
that isospin is conserved, i.e. the relative energies of the proton levels and the 
relative energies of the neutron levels must be the same.  That is, the proton levels can only be shifted by
a constant energy relative to the neutron levels. The experimental studies of nuclei belonging to an 
isospin multiplet shows that the relative energies of exited states agree with each other within about $100 keV$.
These Coulomb shifts are not properly accounted for by the differences between the proton and neutron single particle energies
of the Nilsson model (or other potentials). For this reason, we chose to take the average. An overall shift of the proton levels by a constant energy 
results in a constant shift of the average single particle energy, which does not matter, because we only consider energy differences.

The bunching of the single particle levels generates the shell effects 
in the binding energies. In the framework of the Micro-Macro method the
shell correction is the sum of the single particle energies of all occupied
levels minus the Strutinsky average of this sum. The latter term 
is a smooth function of $N$ and $Z$. For the
energy  differences investigated in this paper, it either (nearly) cancels out
or can be considered being  incorporated in the  $CT(T+1)$ term of the model (cf. section \ref{sec:theta}). 
Thus  adding the sum of the single particle energies below the seven-level window
will correctly reproduce the shell correction to the energy differences.

The model contains two parameters, strength of the isovector pair interaction   $G_V$, 
and the parameter $C$ of the \textquotedblleft symmetry\textquotedblright  interaction, which have been determined by 
simultaneously by fitting the even-even odd-odd mass differences and  the energy difference between 
the $T=0$ and $T=1$ states in the odd-odd nuclei.  

\begin{figure}
  \begin{center}
    \includegraphics[width=8.7cm]{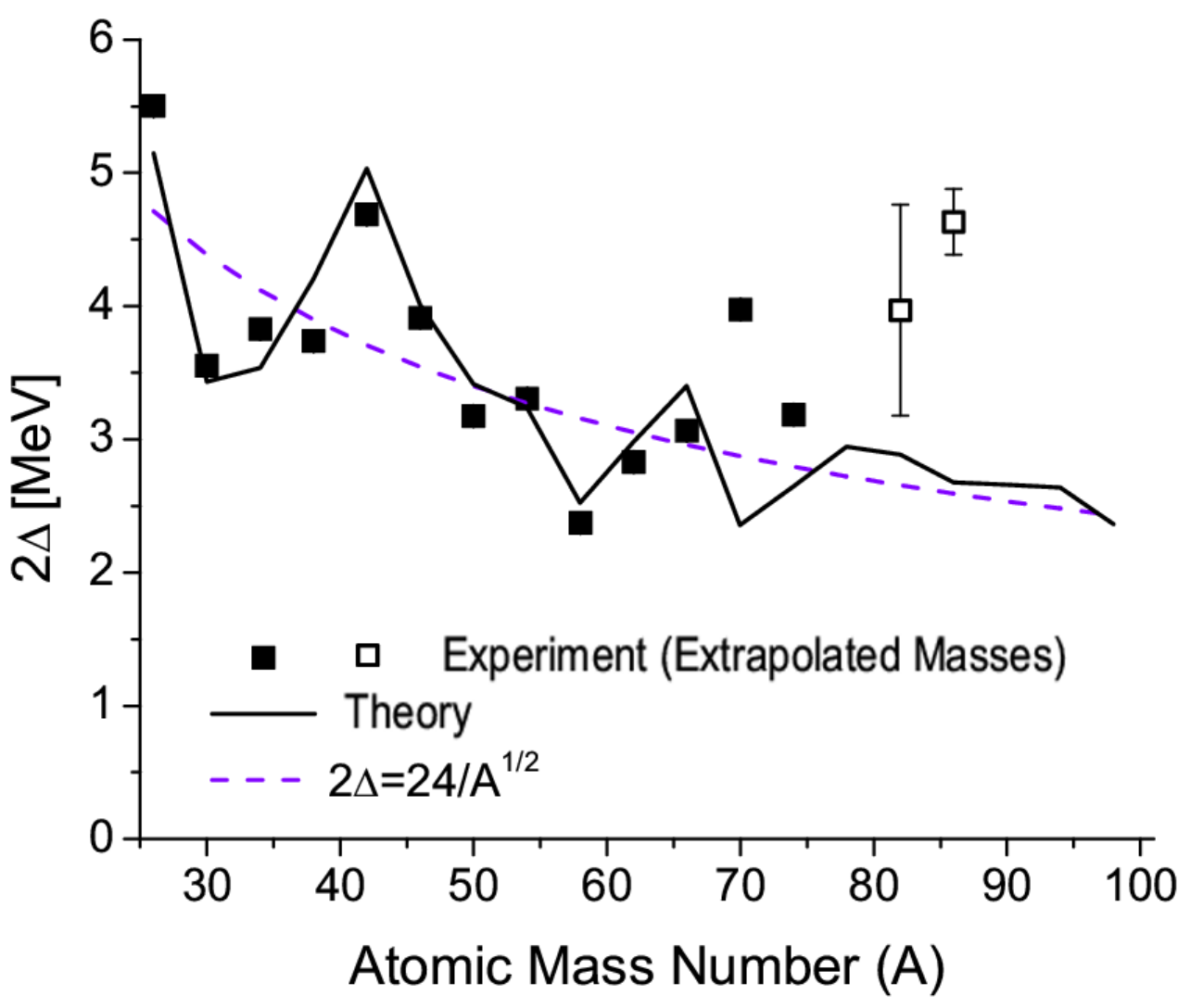}
    \caption{(Color on-line) The even-even odd-odd mass difference $2\Delta$  obtained from (\ref{DeltaEE-OO}) using modified
energies from \citep{Au12} with the Coulomb energy removed. The  solid line shows the calculations.
 The purple dashed line is the global fit of $2\Delta=24 A^{-1/2} [MeV]$. If not visible, the error bars are smaller than the size of the symbols.} 
    \label{fig:EEOO}
  \end{center}
\end{figure}

As well known from 
BCS theory, the value of $G_V$ has to be adjusted to the number of levels taken into account. 
We adopt the standard procedure to reproduce the experimental values of the even-odd
mass differences, which scatter around the smooth  
dependence on the atomic mass number of:
\begin{equation}
\label{DeltaofA}
\Delta \approx \frac{12}{A^{1/2}} [MeV],
\end{equation}
for both protons and neutrons (e.g.\citep{MN92},\citep{NR95}). 
Since our computer code can only handle even-$A$ nuclei, we use the 
the mass differences between the even-even and odd-odd   $N=Z$ nuclides 
derived means of the 3-point formula:
 \small
\begin{equation}
 \label{DeltaEE-OO}
2\Delta(N,Z)=\frac{B_E\textrm{(N-1,Z-1)-2}B_E\textrm{(N,Z)+}B_E\textrm{(N+1,Z+1)}}{2},
\end{equation}
\normalsize
to determine $G_V(A)$.  Coulomb, surface, volume, and symmetry terms in the binding energy
approximately cancel out using this difference. The mass differences have a global dependence on $A$
roughly twice that given by Eqn. (\ref{DeltaofA})  \citep{MF00}.
Eqn. (\ref{DeltaEE-OO}) was evaluated using the binding energies of the even $N=Z$ nuclides
and   the binding energies of the $T=0$ states of the odd $N=Z$ nuclides. 
The following fit was adopted:
\begin{equation}
\label{GofA}
G_V=\frac{13.9}{A^{3/4}} [MeV].
\end{equation}
The values based on experimental binding energies were compared with the ones 
obtained from calculated energies  using $G_V(A)$ as given by Eqn. (\ref{GofA}) and Nilsson levels corresponding to 
AutoTAC equilibrium deformation. In calculating  the odd-odd nuclei,  the fourth (middle) 
level was blocked, because the $T=0$ states have two quasi-particle character with respect to isovector pair correlations.
The blocking procedure is described in detail in
\citep{RS80}.  In essence, the blocked level was disregarded in constructing the matrix and twice its energy was added
after the diagonalization.  Figure \ref{fig:EEOO} compares the experimental with the calculated values. Overall, there is good agreement.
 The deviations are
likely a result of inaccuracies of the Nilsson levels. The deformations resulting from the AutoTAC calculation are often substantially 
smaller than those determined experimentally using $B(E2)$ values.

The competition of the first $T=0$ and first $T=1$ states of odd-odd $N=Z$ was then used  to fix the
 parameter $C$. The theoretical energy difference  $E(T=1)-E(T=0)$ was
 obtained by a seven-level calculation for the $T=1$ state and  for the $T=0$ state by the \textquotedblleft blocked\textquotedblright calculation
 described in the preceding paragraph.    
Without the symmetry interaction term, the fully correlated 
$T=1$ state lies at least  $2\Delta$ below  the blocked $T=0$ state. However, the inclusion of the
symmetry interaction term ($2C$ is added  to the $T=1$ state) results in comparable 
energies for the two states. With increasing $A$ and the levels 
switch order, which is seen experimentally \citep{Tu09} (as
discussed in \citep{JO05}).

Requiring the smooth $1/A$-dependence of the symmetry energy, 
the fit of the calculated  differences $E(T=1)-E(T=0)$ 
to the experimental ones gave:
\begin{equation}
\label{CofA}
C=\frac{58.9}{A} [MeV].
\end{equation}

As already discussed,  odd-odd $N=Z$ nuclei with $A>40$  have a ground state that
has $T=1>T_z=0$.
The inversion of the isospin order has been explained by Refs. \citep{Vo00} and
\citep{MF00}. The $T=0$ state in the odd-odd nucleus is  lifted relative to the
$T=0$ ground state of the even-even neighbors, by
the two quasi-particle excitation energy $2\Delta$. The $T=1$ state
is lifted by the  isorotational  energy $1/\Theta$, which is somewhat smaller 
than $2\Delta$.

\begin{figure}
  \begin{center}
    \includegraphics[width=8.7cm]{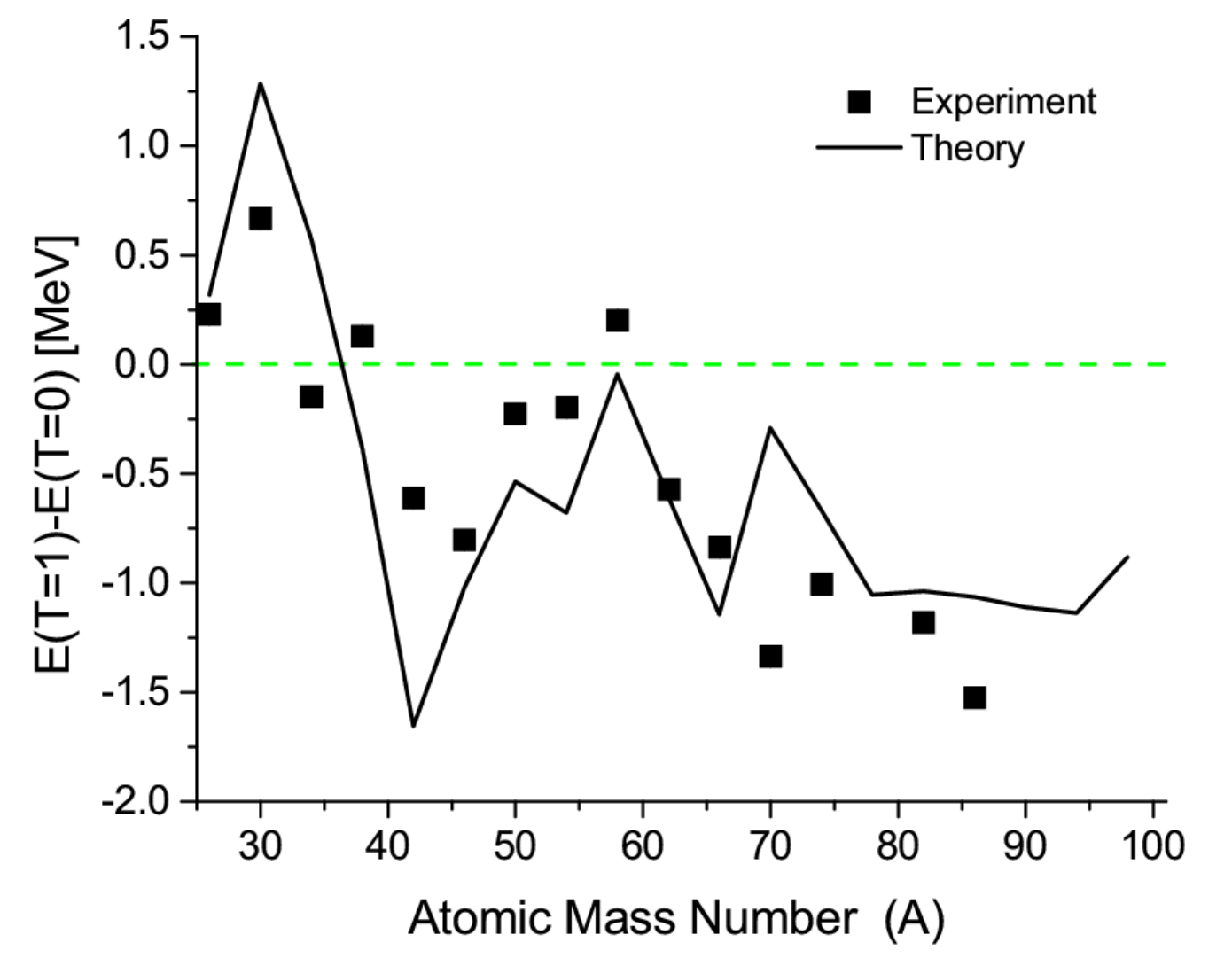}
    \caption{(Color on-line) Energy difference of the first $T=1$ from the first
$T=0$
states in odd-odd N=Z nuclei. The  solid line shows the calculations. Experimental
data from NNDC \citep{Tu09}. The green dashed line indicates where inversion occurs.} 
    \label{fig:inversion}
  \end{center}
\end{figure}

Figure \ref{fig:inversion} shows that calculations  well reproduce energy
difference between the lowest $T=1$
and $T=0$ states, which measures  to relative strength of the  isovector pair
correlation
and the isorotational energies. 
There are large fluctuations in the theoretical calculations caused by the uneven level spacings, which are
roughly reproduced. The deviations are of the same order as the ones of
$2\Delta$ and have the same origin. 

The coefficient of the symmetry energy in the liquid drop model $4a_A=1/2\theta_{LD}=100 MeV/A$.
The value of $C=58.9MeV/A$ is  consistent with the general estimates \cite{BM99} that 
the interaction part of the symmetry energy amounts to about 
50\% of its total value. 

\section{Results - pure isovector pairing}\label{sec:res} 

\subsection{The isorotational moment of inertia }\label{sec:theta}

\begin{figure}
  \begin{center}
\includegraphics[width=8.9cm]{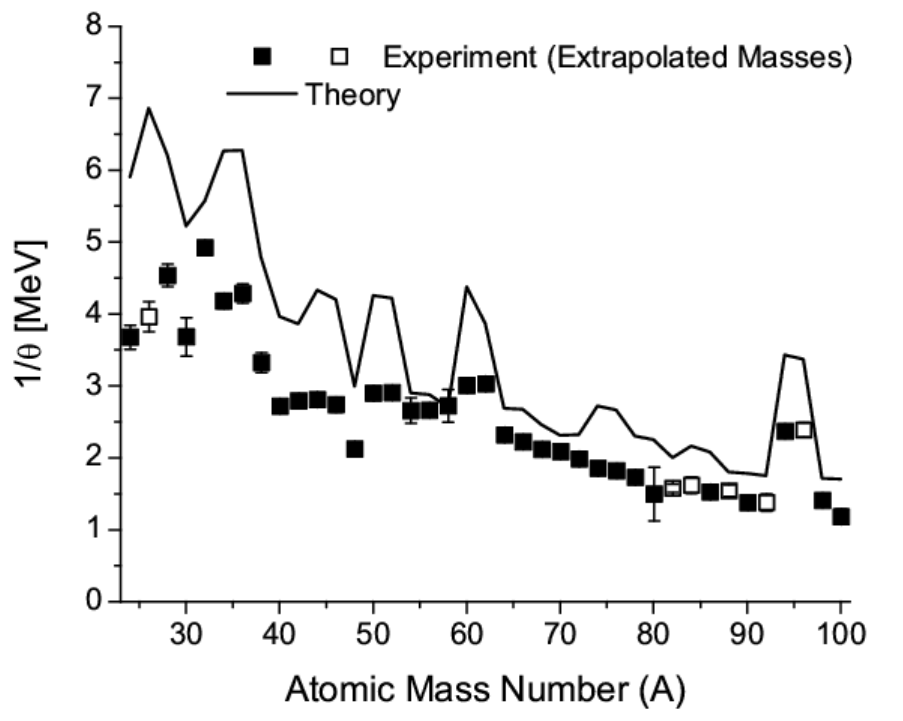}
\caption{(Color on-line) Slope of the isobar energies obtained from (\ref{oneovertheta}) using modified
energies from \citep{Au12} with the Coulomb energy removed. If not visible, the error bars are smaller than the size of the symbols.
The  solid line shows the calculations. }
    \label{fig:slope}
  \end{center}
\end{figure}

Combining the energy differences as:
\footnotesize
\begin{equation}
\label{oneovertheta}
\frac{1}{\theta}=\frac{E_S(T_z=0,1)-2E_S(T_z=2,3)+E_S(T_z=4,5)}{4},
\end{equation}
\normalsize
the local slope for both the even and odd chains of $T_z$ was calculated. 
It has the meaning of the inverse moment of inertia of the isorotational sequence. 
Figure \ref{fig:slope} displays a comparison between experiment and the
theoretical calculation of the slope.

The calculated values of $1/\theta$ are systematically somewhat larger than the
experimental ones.  
We believe that this reflects a fringe effect of our small single particle space.
The number of configurations decreases from 3647 to 1001 and finally to 70 for  
$T=0$, 2, 4 respectively. This results in a decrease of the pair correlation energy, which is
reflected by an increase of $1/\theta$. The consequences of the small number of single particle levels will be discussed 
in more detail in Section \ref{sec:FR}.

\begin{figure}
  \begin{center}
    \includegraphics[width=8.8cm]{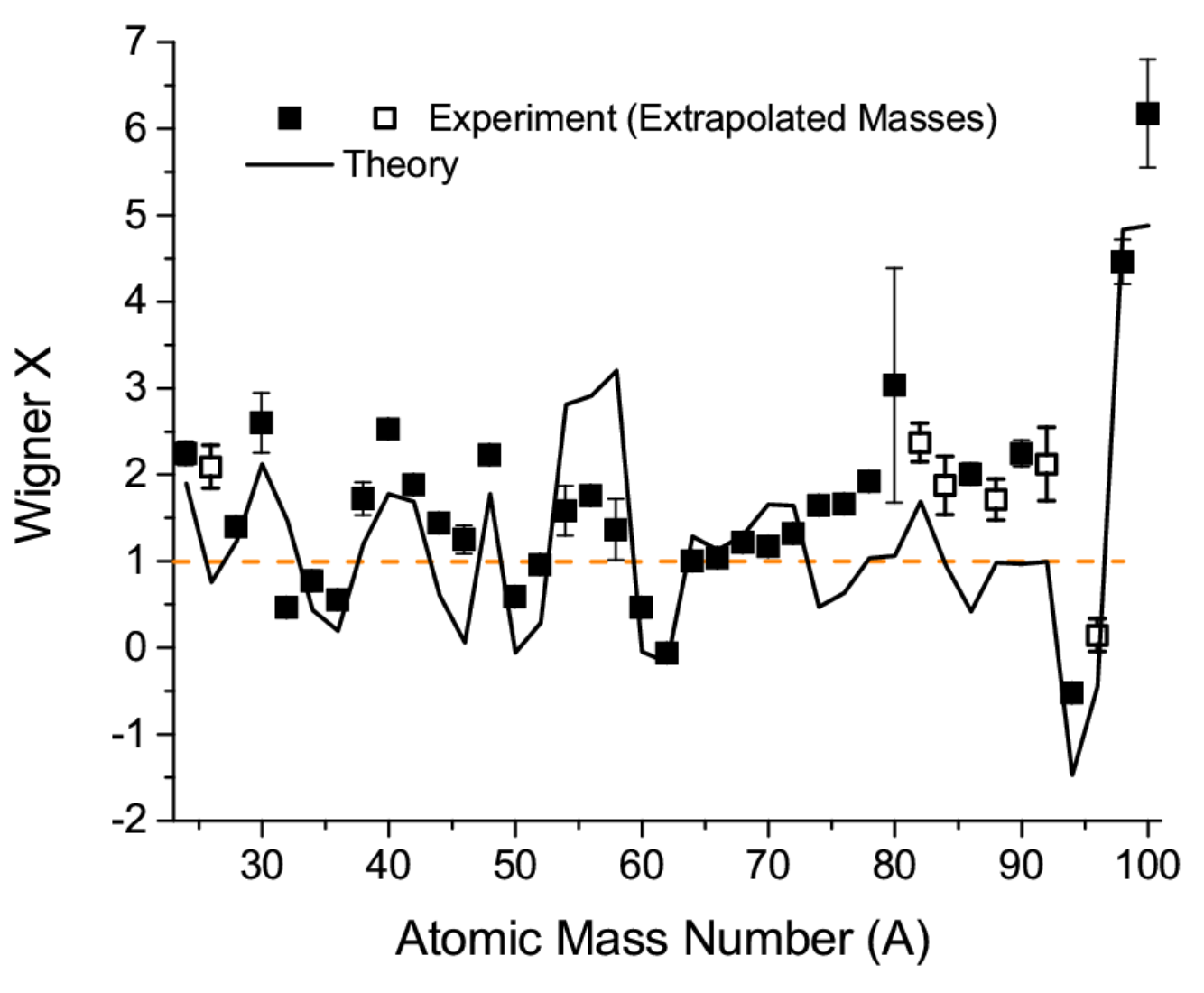}
    \caption{(Color on-line) The Wigner X obtained from (\ref{eqn:XE}) and (\ref{eqn:XO}  
using modified energies from \citep{Au12} with the Coulomb energy removed. If not visible, the error bars are smaller than the size of the symbols.
  solid line shows the calculations. The orange dashed line indicates $X=1$.} 
    \label{fig:Fig4.13}
  \end{center}
\end{figure}

 \subsection{The Wigner X}\label{sec:X}
 
 Figure \ref{fig:Fig4.13} demonstrates that the calculations reproduce well the observed
values of $X$, both the average, which is somewhat larger than 1, and
 the pronounced fluctuations. 
  As seen in Figure \ref{fig:Fig3.12}, 
 the $X$ values derived from the most recent mass tables agree much better with
the calculation than the ones derived by J\"anecke at al. \cite{Ja05} from 
 the 2002 mass tabulations. 
 In the region $A\approx58$, the $X$ values based on the
AutoTAC deformations overestimate the amplitude of the oscillation.
 As shown in the schematic
calculations discussed below, the amplitude of the fluctuations is largest for
strong bunching of levels that occurs near a doubly magic
nucleus. The static AutoTAC  deformations  in this region are mostly zero. 
The degeneracy of the spherical levels will be partially lifted by
shape vibrations, which will damp the fluctuations.   

 There is a tendency that the calculations
underestimate $X$ for $74\leq A \leq 92$. There are two possible justifications for the 
discrepancies seen in this region.
 The experimental uncertainties are large in this mass region and several
binding energies are extrapolated.
Additionally, the AutoTAC  deformations are moderate and fairly constant,
which results in $X$ values close to one. 
Experimental yrast energies of these nuclei indicate a change  
from more vibrational to more rotational behavior as $T_z$ increases, which should be caused by an increasing deformation.  

\begin{figure}[t]
  \begin{center}
  \includegraphics[width=8.7cm]{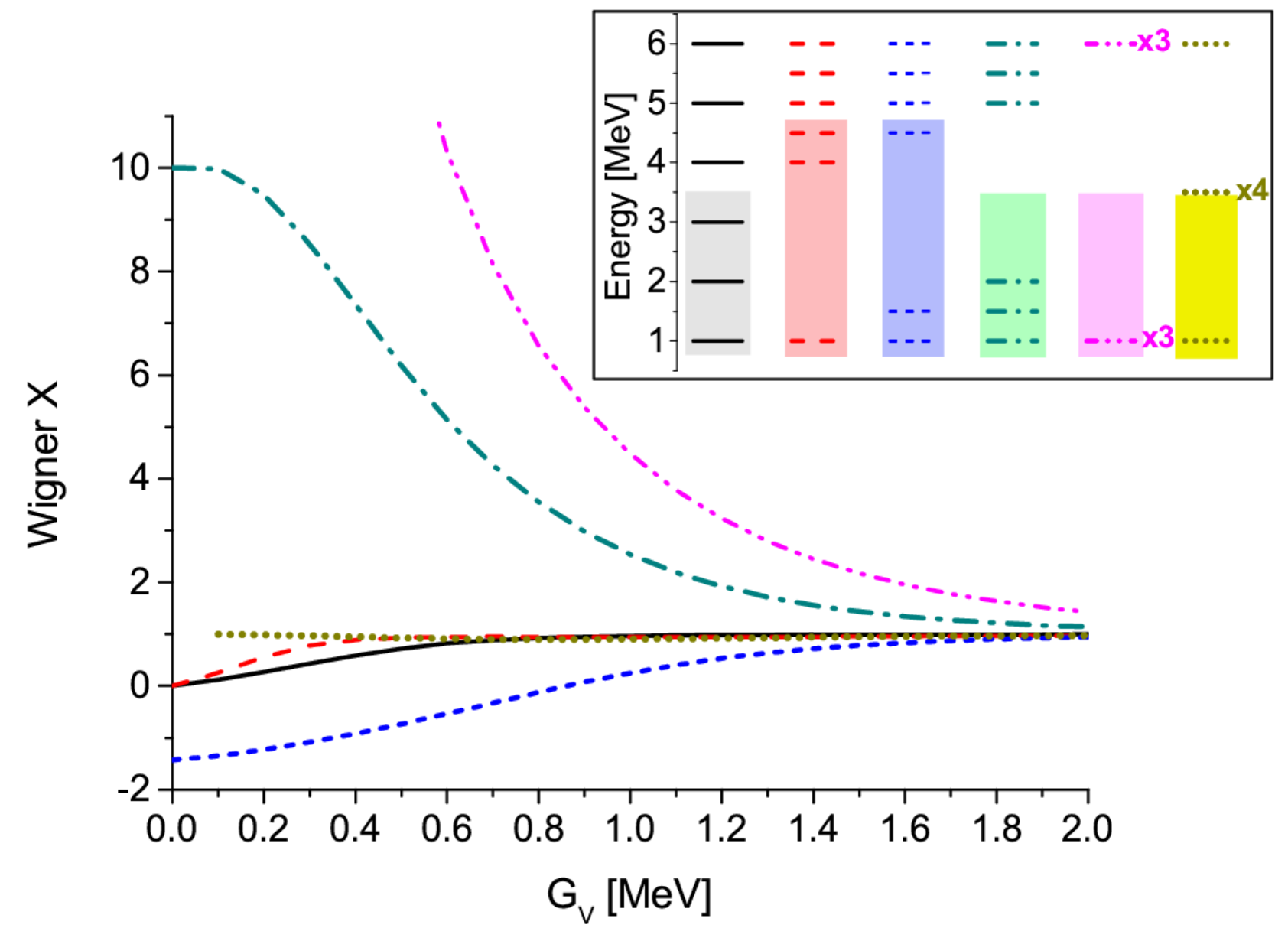}
    \caption{(Color on-line) Wigner X for various level arrangements.
For $G=0$ and $T_z=0$ , the levels 1,2,3 are occupied and 4,5,6 empty. 
Configurations with larger $T_z$ are generated by removing  proton pairs 
and placing neutron pairs according to the Pauli Principle. The average spacing of the energy levels is $1 MeV$.
Note $C=0$.}
    \label{fig:scheme1}
  \end{center}
\end{figure}

The fluctuations of the calculated quantities 
reflect the irregular level spacings.
Figure \ref{fig:scheme1} illustrates the effects on the observed $X$ caused by
changes in level density. The system with even level
spacing is intended to simulate well deformed nuclei. The gaps in the spectrum
simulate the bunching of levels for nuclei with a nearly spherical shape.
In the strong pairing limit, the isorotational band structure is restored and
$X=1$. To approach the limit, the pair field  $\Delta$ must be several times the
average level spacing,
such that local fluctuations of the latter are averaged. The  interaction
strength   $G_V < 1.5 MeV$  is not strong enough to averaging out
fluctuations of $X$ about $1$. The same holds for the fluctuations of $2\Delta$ and $1/\theta$. 

\begin{figure}
  \begin{center}
    \includegraphics[width=8.7cm]{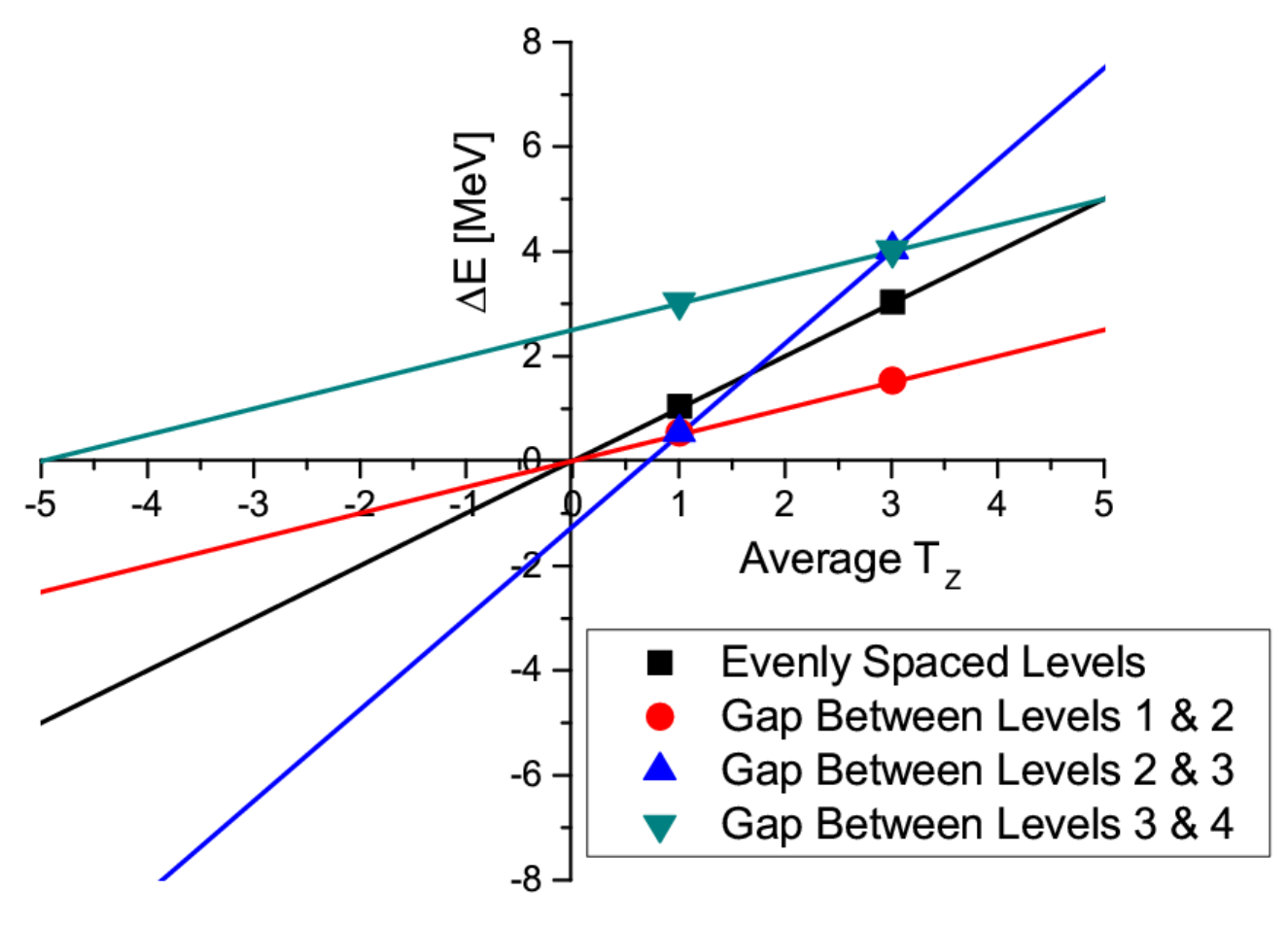}
    \caption{Linear fits corresponding to energy differences for the levels shown in Figure \ref{fig:scheme1} without pair correlations. The 
    slope and intercept are related to $1/ \theta$ and $X$, respectively. }
    \label{fig:scheme2}
  \end{center}
\end{figure}

The deviations from the smooth trends can be understood by considering the
 limit $G_V=0$, when 
the energy is simply the sum of the energies of the occupied levels.
Figure \ref{fig:scheme2} illustrates that the various level
distributions generate different values of $X$ and $1/\theta$ for $G_V=0$, which are still apparent in
the calculations with realistic $G_V$ values.
 In particular, the strong up-down of $X$, which is seen experimentally around
$A=40,\ 56$ and possibly $A=100$,
is caused by  moving through the respective shell gaps.  

\section{Influence of the  Isoscalar Interaction}\label{sec:IVIS}

We have also studied the possible influence of isoscalar proton-neutron pair correlation
on the Wigner $X$   by supplementing  the Hamiltonian  (\ref{eqn:IVPH}) with the term:
\begin{equation}
\label{IVSPH}
 H_{V+S}=H_V- G_S
\sum_{kk'} \hat{S}^+_{k}  \hat{S}_{k'}, ~~~ \hat{S}^+_{k} =\frac{1}{\sqrt{2}}\Big(\hat{n}^+_{k}\hat{p}^+_{\bar{k}}-\hat{p}
^+_{k}\hat{n}^+_{\bar{k}}\Big),
\end{equation}
where $G_S$ is the isoscalar interaction strength. 
The isoscalar pair operators  create  proton-neutron pairs in states with opposite 
projection of the angular momentum. The rational for using such an interaction is that
the strong spin-orbit coupling generates  this type of degenerate time reversed states, which 
are expected to be correlated.  This has been used by Chasman before \citep{Chasman2002}, and the inclusion of this type of interaction 
into our model is straight forward, however it generates
 many more configurations. For example, a six level calculation using  the pure isovector Hamiltonian (\ref{eqn:IVPH}) 
 with six protons and six neutrons has 1001 configurations, 
 while including the isoscalar contributions results in 1992 configurations. As a result, only six levels could  be used for the 
 isovector plus isoscalar calculations. 
 The dimensions are 1992, 825, 66 for $T$=0, 2, 4, respectively.
\begin{figure*}
  \begin{center}
    \includegraphics[width=17cm]{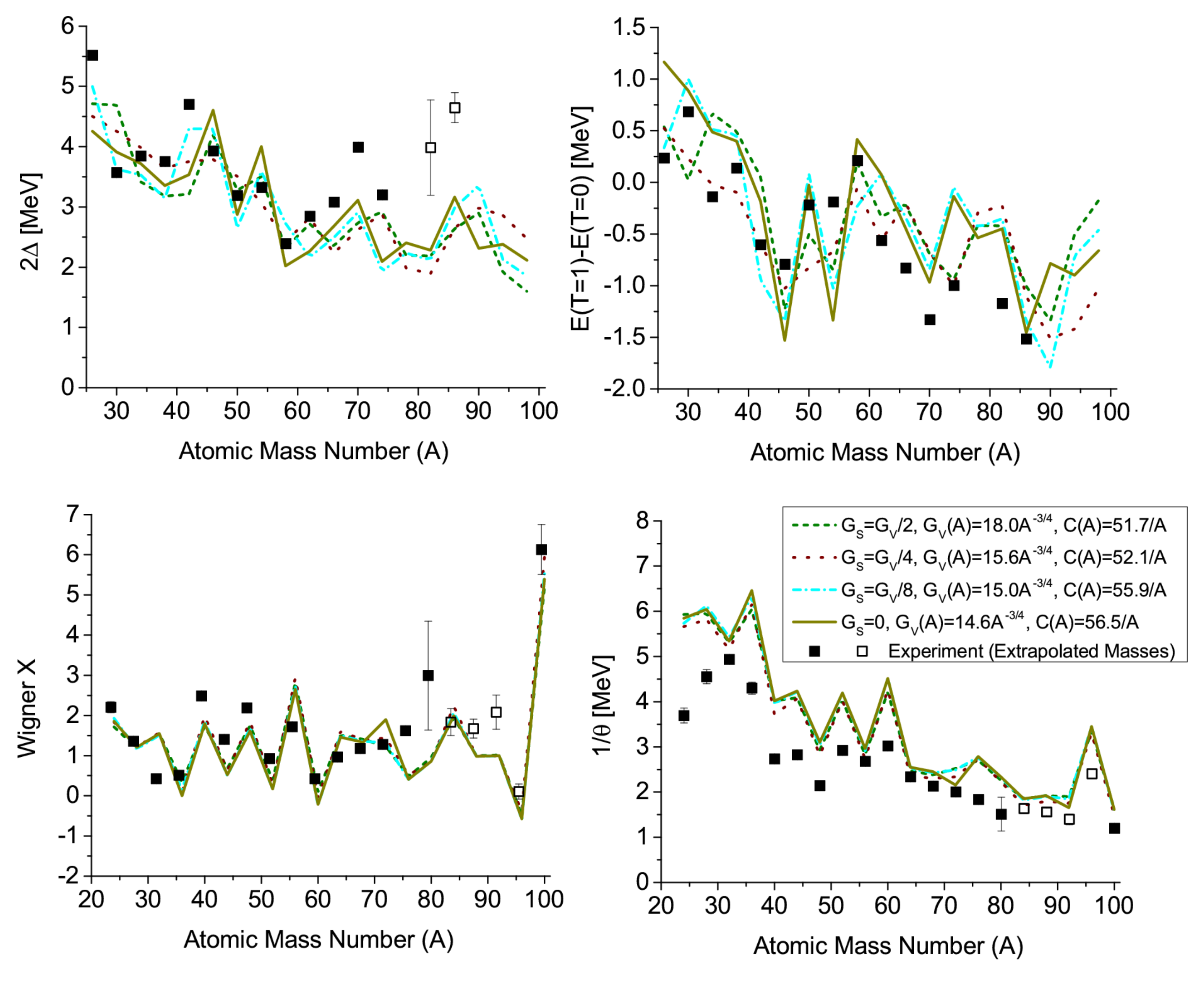}
    \caption{(Color on-line) $2\Delta$, $E(T=1)-E(T=0)$, $X_E$, and $\frac{1}{\theta}$, 
    using the fixed interaction strength ratios, and listed of $G_V(A)$, and
$C(A)$ (both in units of $MeV$), compared to experiment. If not visible, the error bars are smaller than the size of the symbols.
    } 
    \label{fig:GRatio}
  \end{center}
\end{figure*}

The calculations were carried out for fixed ratios of $G_S/G_V=[0,\frac{1}{8},\frac{1}{4},\frac{1}{2},1,2,4,8]$, using the six Nilsson levels 
nearest the Fermi surface, which were determined as described above. 
The parameters  $G_V(A)$ and $C(A)$ were determined as before by fitting the experimental values of $2\Delta$ and $E(T=1)-E(T=0)$.
The results are summarized in Figure \ref{fig:GRatio}. 
 Note that the $G_S=0$
calculation differs from the previously discussed calculations because the number of levels has changed. This reduction of the 
number of levels from seven to six requires a renormalization of $G_V$, which increased by about $5\%$. 
The  $C(A)$ values are also renormalized and decreased by about $4\%$.
The actual fit values used are included in the figure. As a whole, the results
only insignificantly change within the displayed range of the ratio between the interactions.
For $G_S>G_V/2$ it was not possible to simultaneously fit $2\Delta$ and $E(T=1)-E(T=0)$.
 
Figure \ref{fig:IVIS} shows schematic calculations of the four quantities of interest for six equidistant levels.
The value of $C=1 MeV$ was chosen with the intention to simulate nuclei near $A=60$. As expected, for $G_S$ substantially larger than $G_V$, the 
even-even odd-odd mass differences approach zero and become slightly negative.
This is the signature of an isoscalar pair condensate, where even-even and odd-odd $N=Z$ nuclides merge into 
a pair-rotational band \citep{FS00,FR01}. In order to remain within the experimental band of $2\Delta$, the isovector
correlations must prevail. That is,  the stripe of experimental values of $2\Delta$ lies always  below the diagonal. 
Note that the value of $2\Delta$ does not depend on $C$, because it involves a comparison of $T=0$ states only.  

Comparing the four panels in Fig. \ref{fig:IVIS}, one notices that the quantities  $E(T=1)-E(T=0)$, $X$ and $1/\theta$ 
do not change much along a contour  of constant $2\Delta$, as long as one stays within the band of experimental values delineated by the dashed lines and  
the interval $0<G_S/G_V<0.5$, which is the range of ratios shown in Figure \ref{fig:GRatio}.
This helps to explain why the experimental data could be equally well reproduced within this range. Hence, the coexistence of a moderate
amount of isoscalar pair correlations  is consistent with the data, which however does not provide evidence for its existence.
Large scale Shell Model Calculations with realistic effective interactions find moderate isoscalar pair correlations coexisting
with strong  of isovector correlations  \citep{Poves1998203,Langanke1997253}. 

Our  results do not concur with  Refs. \citep{SW97,Brenner19901,Chou1991487,Satula1997103}  who
relate  the Wigner energy to the presence of  isoscalar proton-neutron pair
correlations. However, they are consistent with the findings of 
Ref. \citep{Poves1998203}, who pointed out  that although the Wigner term is
related to  the $T=0$ part of the residual interaction
in shell model calculations this does not necessarily imply that it is generated by   
proton-neutron pair correlations.   

\section{Fringe effects}\label{sec:FR}
The small number of single particle levels among which the pair correlations are allowed to act causes artifacts  that
will be quantified now. 
As discussed in section \ref{sec:theta}, the number of configurations available for pair correlations strongly decreases with  $T$ (by a factor of 50)
when the Fermi level approaches the upper single particle level, because  the combinatorial possibilities are  reduced. 
This results in an artificial reduction of the pair correlations, which we called the fringe effect. The reduction of the pair correlation energy
increases $1/\theta$, which is in our view the reason why this quantity comes out  systematically 
somewhat  too large.

\begin{figure*}
  \begin{center}
    \includegraphics[width=17cm]{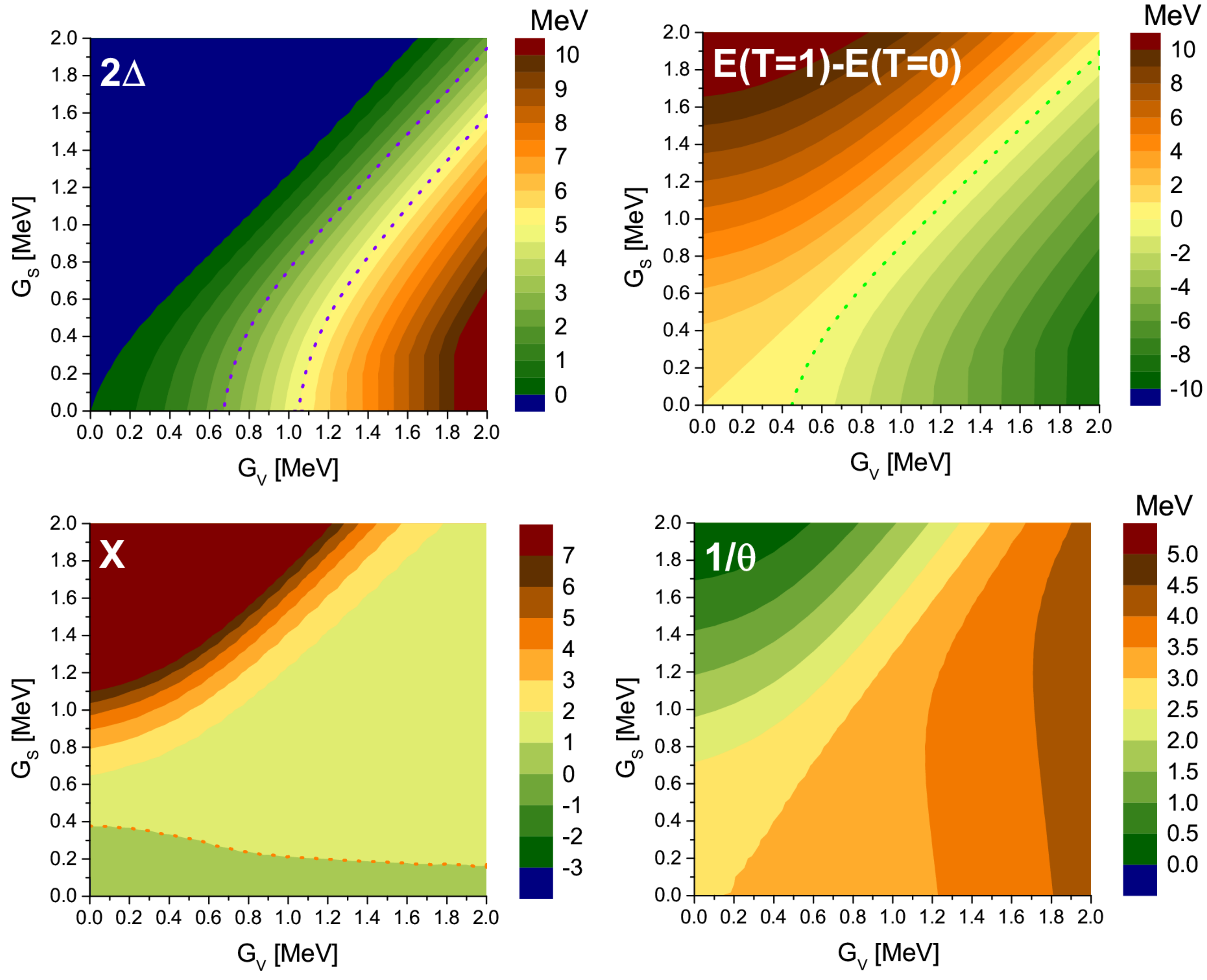}
    \caption{(Color on-line) Isovector plus  isoscalar calculation for
equidistant levels.
    The dotted purple lines indicate  the smoothed trends of $2\Delta$ in 
    Figure \ref{fig:EEOO} around  $A=20$ and $A=100$, respectively, the dotted
    green line $E(T=1)=E(T=0)$, and the dotted orange line $X=1$. The  level spacing  is $1 MeV$ and $C=1MeV$.}
    \label{fig:IVIS}
  \end{center}
\end{figure*}

The fringe effect on $1/\theta$ can be estimated on the basis of  Figure \ref{fig:IVIS}, which shows calculations for six equidistant levels.
The slope $1/\theta$  increases  from 3.00 $MeV$ to 3.32 $MeV$ when $G_V$ changes from 0 to 
a realistic value of 1 $MeV$,  while holding $G_S=0$. This puts a scale on the fringe effects, because  $1/\theta$ should not 
change with the strength of the pair correlations for a sufficiently large set of equidistant levels. Instead, it should stay equal to the value
without pair correlations. More specifically, it should be $d+2C$, with $d=1 MeV$ being the average level spacing and $C=1 MeV$ being the strength of the symmetry interaction for this case.
The difference of 0.32 $MeV$ is consistent with the systematic overestimation of $1/\theta$ in Figs. \ref{fig:slope} and \ref{fig:GRatio} 
around $A=80$. 

The six-level and seven-level calculations give nearly the same values of $1/\theta$. This can be seen by comparing Figs. \ref{fig:slope} with
\ref{fig:IVIS}. Note, in the latter only even $T$ chains have been evaluated. More quantitatively, the respective mean values of $1/\theta$ are 3.49 $MeV$ and 3.47 $MeV$, and
the respective mean square deviations are  0.90 $MeV$ and 0.93 $MeV$. The contribution of the symmetry interaction to  $1/\theta$ is equal to $2C$. The fact that 
a smaller $C$ in six-level calculation gives the same values of $1/\theta$  as the seven-levels calculation means that the fringe effect must be larger for six than for seven levels,
which is compensated by the reduction of $C$. For $A=80$ the difference of $2C$ between the seven and six level cases is 0.06 $MeV$. This increase 
of $2C$ compensates
a decrease of the fringe effect in the seven-level calculations by the same amount. This value is 20\% of the 0.32 $MeV$ estimate in the preceding paragraph.

 Using approximately
$60\%$ the value for $C(A)$ would reconcile the discrepancy in $1/\theta$ between experiment
and theory. However it would result in a systematic overestimate of $E(T=1)-E(T=0)$.
This fringe effect, which is a limitation of our few-level approach, lead us to
use $E(T=1)-E(T=0)$ in the odd-odd $N=Z$ to adjust the $C$ parameter 
nuclei, where the effect is weakest, instead of determining it from the experimental slope of the symmetry energy.
The main focus of our work is the study of the Wigner X, which impacts the nuclei near $N=Z$
strongest. The studies of  Refs. \cite{Ja05} and \cite{MF00} demonstrated that the 
experimental values of $1/\theta$ and $2\Delta$ are consistent with $E(T=1)-E(T=0)=2\Delta-1/\theta$ on average.
We expect that including more single particle levels into the beyond-mean field description of the pair correlations
will resolve the modest inconsistency. Unfortunately, direct diagonalization of the
pairing Hamiltonian will not be feasible because of the combinatorial explosion of the dimensions. A shift of the single particle window
to have the same number of levels on both sides of the Fermi level violates isospin conservation, which is a crucial ingredient.
Clearly   one has to employ some approximation scheme that ensures good isospin. Work along this line is on the way.

\section{Conclusions} 

We have demonstrated that a model based on single particle levels in a deformed
potential,
isospin conserving isovector monopole pairing, and a schematic \textquotedblleft symmetry\textquotedblright interaction
proportional to $\vec T^2$ 
reproduces the term linear in $|N-Z|$ in the nuclear binding
energy. The pairing correlations were treated exactly
by numerical diagonalization in a space of seven single particle levels, which
ensured that isospin was conserved. Isospin invariance requires the coupling constants of the proton-proton, proton-neutron, and
neutron-neutron interaction to be equal. 

The Wigner term appears as a result of breaking   
isospin invariance on the mean field level. The deformation in isospace gives rise to 
an isorotational band with energies $\propto T(T+1)$.  The deformation is caused by the 
isovector pair field and the differences between the proton and neutron nuclear potentials to about equal parts.
 
The model does not introduce new parameters as compared to standard mean field
approaches. The two model parameters are
the pairing strength, which  is fixed by the even-even to odd-odd mass difference, and 
the strength of the symmetry interaction, which is determined by the energy difference between the lowest $T=0$ and $T=1$ states in
odd-odd $N=Z$ nuclei. Using this approach it is possible to get roughly the correct 
order ($T=0$ below $T=1$ for $A<40$ and $T=1$ below $T=0$ for $A>40$).

Merging the symmetry term and the Wigner term of the binding energy  
into one expression of the form \mbox{$T(T+X)/2\theta$}, the values of $X$ are
found to scatter around 1.
The limit $X=1$ corresponds to a regular isorotational band, which emerges if 
isospin is strongly broken by the pair field. Because the realistic pair field has
only moderate strength
 the bunching of the single particle levels, resulting from shell structure, causes strong
fluctuations of the Wigner energy
which are fairly well described by the model.  The remaining
deviations can be attributed to inaccuracies of the calculated single particle
energies. 

A combination of an
isorotational invariant effective interaction in the particle-hole channel with
isovector pairing
interaction is capable of reproducing the Wigner energy, provided  the pairing
correlations
are treated beyond the mean field approximation and isospin is conserved. 
How to accomplish this for 
the present standard  mean field approaches remains to be studied. In a future
study we will address this
question by comparing our results  with approximations as e.g. isospin
projected mean field solutions.  

In addition, we  investigated how including a monopole isoscalar pairing
interaction would modify the results. As long as the the ratio
between the isoscalar and isovector coupling constants remained smaller than 0.5,  the experimental
 values of the Wigner energy and
of the $T=0$-$T=1$ energy difference in odd-odd $N=Z$ nuclei could be 
equally well reproduced after a slight readjustment of the two model parameters. The results turned out to be
 insensitive to moderate isoscalar pair correlation of this scale and, thus, did not provide any clue
 about their possible presence. Ratios of the isoscalar-isovector coupling constants larger than 0.8 
  contradict  the experimental values of the even-even odd-mass mass differences.   

Supported by the DoE Grant DE-FG02-95ER4093.

\bibliography{bibliography} 

\end{document}